# AlN/β-Ga$_2$O$_3$ based HEMT: a potential pathway to ultimate high power device

*Yi Lu$^{1,2,*}$, Hsin-Hung Yao$^{1,*}$, Jingtao Li$^{1}$, Jianchang Yan$^{2}$, Junxi Wang$^{2}$, Jinmin Li$^{2}$, Xiaohang Li$^{1,\#}$*

[1] King Abdullah University of Science and Technology (KAUST), Advanced Semiconductor Laboratory, Thuwal 23955-6900, Saudi Arabia

[2] Research and Development Center for Solid State Lighting, Institute of Semiconductors, Chinese Academy of Sciences, Beijing 100083, China

## ABSTRACT

Gallium Oxide (Ga$_2$O$_3$) has a huge potential on the power device for its high breakdown filed and good transport properties. β-Ga$_2$O$_3$ as the thermodynamics stable phase, has been demonstrated to form high electron mobility transistor (HEMT) through δ-doping in the barrier due to its none-polar property. Following the development in III-V HEMT which turns from δ-doping-induced to polarization-induced 2DEG, an alternative method based on III-N materials/β-Ga$_2$O$_3$ heterostructure is proposed that utilizing the polarization difference on the interface. Further requirements of electric field and conduction band difference show that only nitrogen (N)-polar AlN on β-Ga$_2$O$_3$ can form the channel and hold large 2DEG concentration on the interface. Compared with conventional metal-polar AlN/GaN HEMT, the proposed N-polar AlN/β-Ga$_2$O$_3$ HEMT show a much larger 2DEG concentration, the spontaneous-polarization-dominated electric field, better DC output performance, as well as higher breakdown voltage. This study provides a new research approach that shifting from δ-doping-induced to polarization-induced on β-Ga$_2$O$_3$-based HEMT, which can also be a guideline for community excavating the application potential of Ga$_2$O$_3$.

## INTRODUCTION

Gallium(III) Oxide belongs to a family of conducting transparent semiconductor (TSO) with the formula Ga$_2$O$_3$. Recently, Ga$_2$O$_3$ attracts many attentions[1] due to its wide band gap and high breakdown field properties, which has applications in deep UV[2] and power electronics[3,4]. Researchers have demonstrated that Ga$_2$O$_3$ power device has critical field strength as high as 3.8 MV/cm[5], and its theoretical critical field strength is 8 MV/cm, making Ga$_2$O$_3$ an ideal candidate for next-generation power devices. In comparison, commercialized SiC and GaN power device have theoretical critical field strength 3 and 3.14 MV/cm, respectively. Ga$_2$O$_3$ has five commonly identified polymorphs, corundum (α), monoclinic (β), cubic (γ), and orthorhombic (ε), and bixbyte (δ).[5,6,7,8,9] Except for β-phase, the rest of the polymorphs are not stable, and have phase transition to β-phase after annealing.[10] Though thermodynamically unstable, α-phase and ε-phase have potential applications due to unique properties. α-phase Ga$_2$O$_3$ and α-phase (AlGa)$_2$O$_3$ alloy on α-phase Al$_2$O$_3$ (sapphire) has been demonstrated by mist-CVD, which has potential for power device application.[11] ε-phase Ga$_2$O$_3$ has the largest polarization constant, which have the potential for polarization engineering.[12] β-Ga$_2$O$_3$ is thermodynamics stable phase, and has attracted most of the recent attention. Its Schottky barrier diode (SBD)[13], enhanced-mode vertical transistor[14], and high electron mobility transistor (HEMT)[15] have been realized based on its good transport properties (mobility>200cm$^2$/Vs, saturation velocity ~2 × 10$^7$cm/s[16]).

Due to its nonpolar property, the demonstrated β-Ga$_2$O$_3$ based HEMT such as β-(Al$_x$Ga$_{1-x}$)$_2$O$_3$/Ga$_2$O$_3$ heterostructure follows conventional GaAs HEMT structure[17] to apply δ-doping in the barrier to induce the 2DEG in channel and the sheet charge density of ~10$^{12}$/cm$^2$ can be achieved.[18] However, compared with doping-induced HEMT, utilizing the polarization property can realize superior performance. A good example is that polarization-induced GaN-based HEMT exhibits higher 2DEG density of the order of ~ 10$^{13}$/cm$^2$ [19] than GaAs-based HEMT with δ-doping-induced 2DEG (~ 10$^{12}$#/cm$^2$).[20] . If β-Ga$_2$O$_3$ HEMT can follow GaN HEMT structure using polarization-induced 2DEG in channel, the performance would be even better than δ-doping-induced 2DEG HEMT. Recently, many studies or proposes to combine Ga$_2$O$_3$ and nitride materials together to form oxide/nitride heterostructure for potential applications.[21,22,23,24] Also, the heterojunction between β-Ga$_2$O$_3$ and III-nitride material has been demonstrated to investigate the band alignment and other properties[25,26]. A large polarization discontinuity on the heterojunction between the nonpolar β-Ga$_2$O$_3$ and III-nitride with polarization, which may give rise to the formation of 2DEG on the interface, even with an unexceptionable device performance.

In this work, we systematically investigate the interface between III-N materials/β-Ga$_2$O$_3$ based on the band alignment and modern polarization theory. By considering about the conduction band and polarization difference, the N-polar AlN/β-Ga$_2$O$_3$ heterojunction is found that can form a channel and hold large 2DEG density on the interface. Compared with conventional metal-polar AlN/GaN HEMT, the proposed N-polar AlN/β-Ga$_2$O$_3$ HEMT using polarization-induced 2DEG can realize much huger 2DEG density (spontaneous-polarization-dominated), better DC and transconductance performance, as well as higher breakdown voltage.

**POLARIZATION THEORY AND STRUCTURE DESIGN**

**A. Polarization theory**

Two-dimensional electron gas (2DEG) is well known as a large number of electrons accumulating in a thin layer which are free to move in two dimensions, but tightly confined in the third. The carriers confined in a potential are quantized and the motion of carriers also exhibit highly quantum behaviour, such as the quantum Hall effect[27]. Normally the 2DEG appears near the interface of heterostructure that can just transport along the interface with high mobility and form the flow of current. These special properties are utmostly taken to design field effect transistors (FET) such as HEMT. Both the modulation doped heterojunction[28] or polarization effect[29] can all induce the gathering of the electrons to form 2DEG channel.

The formation mechanism of polarization-induced HEMT is strongly relied on the electric field created by polarization difference on the interface of heterostructure, which can be well explained by the existence of the surface donor-like states[30,31]. Assuming these donor-like surface states are located quite deep under the Fermi level with a thin barrier. When getting the positive sheet charge on the interface, a positive electric field along the grown direction can be obtained. Under this circumstance, as the width of the barrier increases, the surface donor level will approach the Fermi level till reach it, and then the electron can be transferred from the surface donor state to the conduction band. At the same time, the electric field created by the positive polarization-induced sheet charge on the interface provides the force for these electrons from the surface state being drafted to the heterointerface. Moreover, the discontinuity of conduction band is needed to create a thin channel where the high-density electrons can exist as well as separate the electrons in the channel from their donor atoms which reduces Coulomb scattering and hence increases the mobility of the conducting electrons. Therefore, a large number of electrons can accumulate within a thin channel with high mobility and form 2DEG. Obviously, the formation mechanism of 2DEG in polarization-induced HEMT can be attributed to three conditions: (i) the polarization-induced sheet charge, (ii) the position of Fermi level and surface state, and (iii) the conduction band difference ($\Delta E_c$).

The demonstrated heterojunctions between β-Ga$_2$O$_3$ and III-nitride material have shown that the c-plane III-nitride material such as AlN or GaN could be a better template for the (-201) β-Ga$_2$O$_3$ epilayer growth and vice versa[25,26]. To our knowledge, when forming a heterostructure with two different materials, especially including III-N material, the polarization effect needs to be investigated and considered. The sheet charge density Δσ induced by polarization difference at the heterointerface can be presented by

$$\Delta\sigma = -\Delta P = P_{sp}^{(sub)} - [P_{sp}^{(epi)} + P_{pz}^{(epi)}] \qquad (1)$$

What's more, if people want to design the polarization-induced HEMT based on β-Ga$_2$O$_3$ and III-nitride material, the polarization difference must be carefully considered due to that the only positive electric field along the growth direction induced by polarization can raise the donor level to Fermi level and draft electrons to the channel.

Generally, ZB polarization engineering which takes zincblende (ZB) as reference structure to calculate spontaneous polarization (SP) and polarization difference of heterojunction has been widely used[32]. It shows a good coherence on the calculation of

III-N materials (like AlN, GaN and InN) including semi-polar or none-polar nitride materials[33]. However, aiming at the calculation of c-plane III-N materials, a modern theory of polarization has been recently proposed and verified [34]. This modern polarization model takes hexagonal reference structure (H reference) rather than the ZB structure to calculate the polarization of c-plane III-N materials, and shows the high accuracy on the spontaneous and piezoelectric constants, as well as bound sheet-charge densities and polarization fields. The calculated polarization values of AlN, GaN ,and InN by H reference are listed in Table I, which has been presented by Liu et. al[35]. Moreover, Macciono et. al. [36] has reported that the most stable β-Ga$_2$O$_3$ shows a zero value no matter on spontaneous polarization (SP) properties or piezoelectric (PZ) constants.

Table I. Structural parameters, SP and PZ constants of AlN, GaN, InN, and β-Ga$_2$O$_3$.

| Parameter | Units | AlN | GaN | InN | β-Ga$_2$O$_3$ |
|---|---|---|---|---|---|
| Lattice constant | Å | 3.113 | 3.182 | 3.541 | a=12.23<br>b=3.04<br>c=5.80 |
| SP | C/m$^2$ | 1.333 | 1.339 | 1.164 | 0 |
| e$_{33}$ | C/m$^2$ | 1.642 | 0.615 | 1.058 | 0 |
| e$_{31}$ | C/m$^2$ | -0.669 | -0.358 | -0.549 | 0 |
| C$_{33}$ | - | 373 | 398 | 224 | 330 |
| C$_{13}$ | - | 108 | 106 | 92 | 125 |

The correct implementation of polarization constants in wurtzite materials shows prodigious amendment on the SP and PZ constant. However, it should be noted that the PZ polarization will have different forms when taking H reference polarization model to calculate the heterostructure due to the strain in epitaxial layer builds a bridge for the SP and PZ polarization. When calculating the sheet charge at the heterointerface [Equation (1)], the PZ polarization can be described as follow:

$$P_{pz}^{(epi)} = 2\left[e_{31}^{(epi)} - P_{sp}^{(epi)} - \frac{C_{13}^{(epi)}}{C_{33}^{(epi)}}e_{33}^{(epi)}\right] \times \frac{a^{(epi)}-a^{(sub)}}{a^{(epi)}} \qquad (2)$$

Here, $e_{31}^{(epi)}$ and $e_{33}^{(epi)}$ are the proper PZ constants of the epitaxial layer, $C_{13}^{(epi)}$ and $C_{33}^{(epi)}$ are elastic constants of the epitaxial layer, strain$^{(epi)}= \frac{a^{(epi)}-a^{(sub)}}{a^{(epi)}}$, and $a$ is the lattice constant.

Typical Al(Ga)N/GaN HEMT adopts a thin Al(Ga)N as a barrier, *ie*. metal-polar AlN barrier fully strained on freestanding metal-polar GaN thus forms a heterojunction. The calculation results of polarization by hexagonal reference in Table II shows a similar spontaneous polarization ($P_{sp}$) between AlN (1.333 C/m$^2$) and GaN (1.339 C/m$^2$) while a large piezoelectric polarization $P_{pz}$ (-0.109C/m$^2$) caused by strain in the epitaxial layer. The sheet charge density (Δσ) determined by the total polarization difference on the interface of AlN/GaN is 0.114 C/m$^2$. The positive sheet charge density on the interface will create an electric field in barrier along the growth direction. The energy level difference between the surface donor level and Fermi level decreases with increasing barrier thickness. At a critical thickness, the donor energy reaches the Fermi level. With a large conduction band difference (ΔEc=1.9eV, shown in Fig 1) at the heterojunction, the electric field created by the positive sheet charge on the interface can force the electrons from surface state to accumulate on the interface and form 2DEG in the channel.

Table II. Spontaneous polarization ($P_{sp}$), piezoelectric polarization ($P_{pz}$), and sheet charge density (Δσ) of conventional AlN/GaN interface and (metal-polar/N-polar) III-nitride /β-Ga$_2$O$_3$ interface. Suppose the epitaxial layers are all fully strained on the substrate.

|  | Al(Ga)N/GaN |  | III-N/β-Ga$_2$O$_3$ |  |  |  |  |
| --- | --- | --- | --- | --- | --- | --- | --- |
|  | AlN/GaN |  | AlN/β-Ga$_2$O$_3$ |  | GaN/β-Ga$_2$O$_3$ |  | InN/β-Ga$_2$O$_3$ |
| Polarity of barrier | Al-polar AlN | GaN | Metal-polar | N-polar | Metal-polar | N-polar | Metal-polar | N-polar |
| $P_{sp}$ (C/m$^2$) | 1.333 | 1.339 | 1.333 | -1.333 | 1.339 | -1.339 | 1.164 | -1.164 |
| $P_{pz}$ (C/m$^2$) | -0.109 | 0 | 0.119 | -0.119 | 0.167 | -0.167 | 0.610 | -0.610 |
| Δσ (C/m$^2$) | | 0.114 | -1.452 | 1.452 | -1.506 | 1.506 | -1.774 | 1.774 |
| $\Delta E_c$ (eV) | | 1.9eV | 1.75eV | 1.75eV | -0.15 | -0.15 | -2.55 | -2.55 |

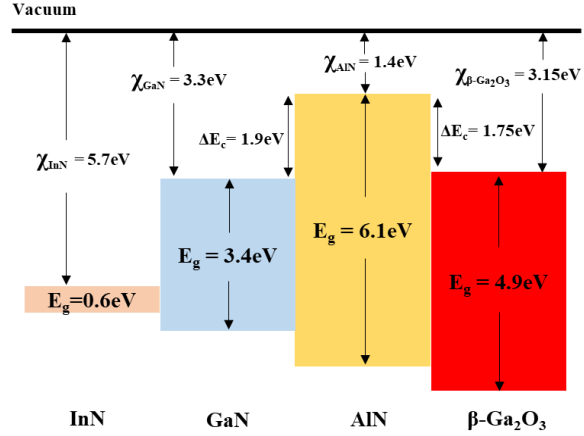

Figure 1. Schematic diagram of the band alignment diagram in the III-N/β-Ga$_2$O$_3$ heterojunction, (Refs. 40,41).

The calculation results by H reference show that the typical metal-polar Al(Ga)N/GaN HEMT requires a positive sheet charge density to generate the electric field for the motion of surface donor level and the drafting of electrons from donor level to the interface. As for the heterojunction of c-plane III-nitride material and β-Ga$_2$O$_3$, the zero-polarization property of β-Ga$_2$O$_3$ results in a large SP discontinuity between the β-Ga$_2$O$_3$ and III-nitride material. When employing the polarization calculation by H reference on the heterojunction, a large sheet charge density can be obtained.

Normally the metal-polar III-nitride material is utilized for their more stable properties and convenient mass production. Besides, N-polar III-nitride material can also be realized through metal organic chemical vapor deposition (MOCVD) or molecular beam epitaxy (MBE) on different substrate and many devices such as LEDs[37] or HEMTs[38] have been demonstrated based on N-polar materials. As a result, the polarization difference of heterojunction which grows metal-polar (or N-polar) III-nitride material (AlN, GaN and InN) on β-Ga$_2$O$_3$ can all be calculated, which is shown in Table II. It can be found that the positive sheet charge density can be gotten for all

N-polar III-nitride material on β-Ga$_2$O$_3$ while negative sheet charge density can be obtained for metal-polar III-nitride material on β-Ga$_2$O$_3$. Compared with the positive sheet charge on the metal-polar AlN/GaN heterojunction, the metal-polar III-nitride material (AlN, GaN, InN) on β-Ga$_2$O$_3$ with negative sheet charge could not meet the demand for the inducing of 2DEG while N-polar III-nitride material (AlN, GaN, InN) on β-Ga$_2$O$_3$ with positive sheet charge (1.452 C/m$^2$, 1.506 C/m$^2$, 1.774 C/m$^2$, respectively) can provide the electric field for the surface donor energy reaching the Fermi level and also for the electrons from the surface state drafting to the heterointerface. According to the band alignment of III-nitride material and β-Ga$_2$O$_3$ shown in Fig 1, the conduction band difference (ΔEc) of III-nitride material (AlN, GaN, InN) on β-Ga$_2$O$_3$ is 1.75eV, -0.15 eV, -2.55 eV, respectively[25,26]. When forming a heterojunction, only the AlN/β-Ga$_2$O$_3$ with ΔEc=1.75eV can provide the bending in conduction band to form a triangle as a channel for the existing of electron. Not only that, AlN/β-Ga$_2$O$_3$ heterojunction with a type-II heterostructure has a much smaller lattice mismatch (around 2.4%) compared with β-Ga$_2$O$_3$/GaN (4.7%)and β-Ga$_2$O$_3$/sapphire (6.6%)[41], which has a huge potential on the development of power devices.

Based on the polarization difference and band alignment of AlN/β-Ga$_2$O$_3$, we can figure out the schematic band diagram Al-polar and N-polar AlN on β-Ga$_2$O$_3$ under a relatively thick barrier (the Fermi level can reach the donor state level) and the schematic diagrams are shown in Fig 2 (a)(b). It is apparent that only positive sheet charge on the interface can induce the accumulation of electron and form 2DEG, which is also consistent with typical AlN/GaN HEMT containing positive sheet charge (0.114C/m$^2$) on the interface. Compared with Al-polar AlN on β-Ga$_2$O$_3$, the Fermi level and the conduction band can form a triangle on the interface that can hold huge amounts of trapped electrons for the heterostructure of N-polar AlN on β-Ga$_2$O$_3$. On the contrary, Al-polar AlN on β-Ga$_2$O$_3$ cannot induce 2DEG with negative sheet charge (-1.452 C/m$^2$) on the interface.

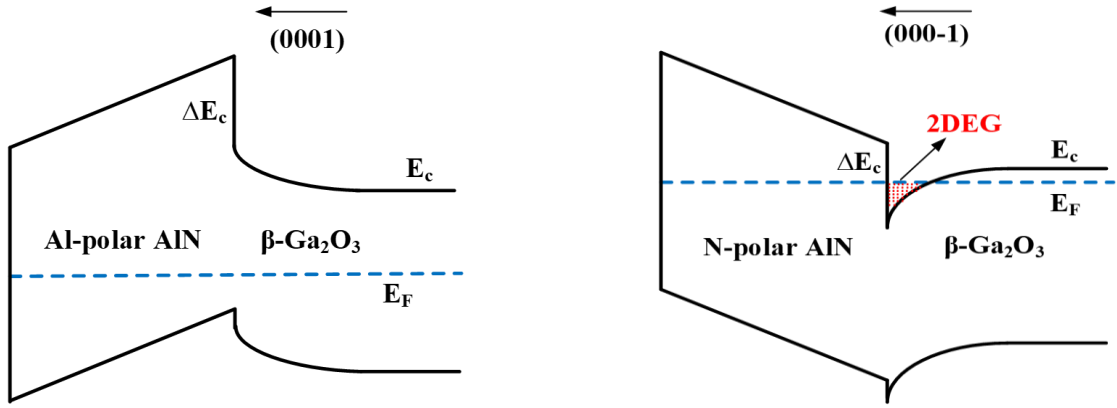

Figure 2. Schematic band diagram of Al-polar AlN (a) and N-polar AlN (b) on β-Ga$_2$O$_3$.

The sheet charge density results are all relatively huge because of the large difference between the SP of III-Nitride and Ga$_2$O$_3$ family. In contrast, piezoelectric polarization's contribution to the interface is relatively minor compared to the spontaneous polarization. This SP-dominant phenomenon shows an abnormal case different from previous theory in III-nitride, where the PZ polarization is often the main concern. At the same time, the heterostructure of III-nitride material on β-Ga$_2$O$_3$ shows a several times larger sheet charge density than wurtzite III-nitride systems usually with an around 0.1 C/m$^2$ sheet charge density and hence a more significant polarization effect. [18] This significant polarization effect could deteriorate the external quantum efficiency of optoelectronics such as LEDs. On the other hand, it is extremely favorable for mass of applications, especially for power device HEMT to have a large amount of 2DEG in the channel.

**B. Device structure design**

The N-polar AlN/β-Ga$_2$O$_3$ heterojunction has been proposed and proved to meet all the standard of forming a HEMT structure, based on which the structure of the proposed N-polar AlN/β-Ga$_2$O$_3$ HEMT and parameter values employed for the simulation are shown in Fig. 3 and Table III, respectively. A 3 μm thick β-Ga$_2$O$_3$ buffer layer where the 2DEG channel exists, is fully relaxed on the 5 μm SiC substrate. The N-polar AlN investigated above is selected as barrier on β-Ga$_2$O$_3$. The dielectric SiO$_2$ material is deposited on top for surface passivation. The length of the Schottky gate (1μm), gate-to-source spacing (2μm) and gate-to-drain spacing (10μm) are all set as typical values of HEMT. In order to suppress the parasitic substrate conduction, traps with a maximum concentration of $1\times10^{14}$ cm$^{-3}$ and a relative energy level of 1.1 eV are also defined as to make the substrate behave as a semi-insulating substrate.

In addition, a conventional metal-polar AlN/GaN HEMT which changes the buffer layer material from β-Ga$_2$O$_3$ to GaN and adopts metal-polar AlN barrier instead of N-polar AlN, while keeps all other parameters setting the same, is designed as a comparison. The barrier for N-polar AlN/β-Ga$_2$O$_3$ HEMT and metal-polar AlN/GaN HEMT are all set as 3nm, which is based on the feasibility of experiment and the lattice mismatch of AlN/β-Ga$_2$O$_3$ and AlN/GaN. The impact ionization of β-Ga$_2$O$_3$.and wurtzite GaN materials are all based on the Chynoweth impact ionization model and Monte Carlo simulation. The electron/hole ionization rates and critical electric fields of β-Ga$_2$O$_3$.and wurtzite GaN can be found in [39,40.] respectively. For the device structure design and simulation, Advanced Physical Models of Semiconductor Devices (APSYS) developed by Crosslight Inc. is employed.[41]

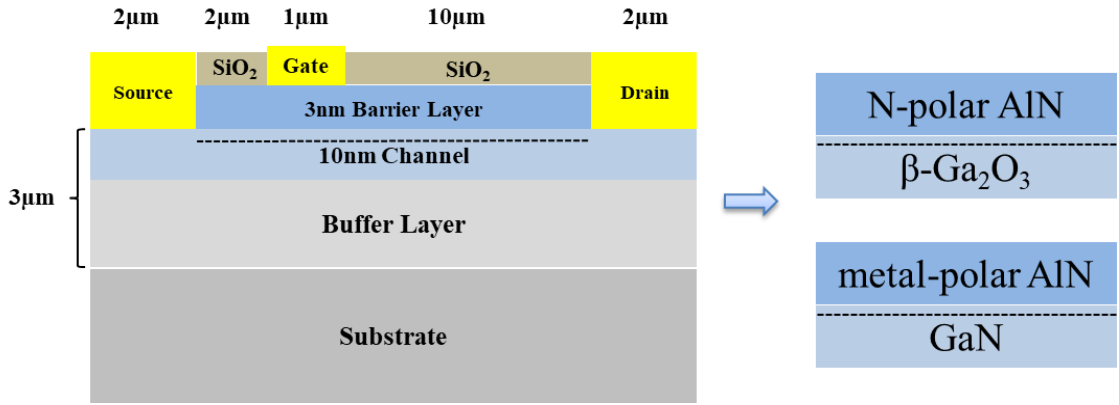

Figure 3 The cross-sectional schematics of the proposed HEMT structure.

Table III. Bandgap, affinity, dielectric constant, electron effective mass, electron mobility, and electron saturation velocity of AlN, GaN, and β-Ga2O3 at room temperature.

| Parameter | Units | AlN | GaN | β-Ga$_2$O$_3$ |
|---|---|---|---|---|
| Bandgap | eV | 6.1 | 3.4 | 4.9 |
| Affinity | eV | 1.4 | 3.3 | 3.15 |
| Dielectric constant | $\varepsilon_0$ | 8.5 | 10.4 | 10.2[42] |
| Electron effective mass | $m_0$ | 0.33 | 0.2 | 0.227[43] |
| Electron mobility | cm$^2$/(V.s) | - | 1200 | 180[44] |
| Electron saturation velocity | m/s | - | $1.91\times10^5$ | $2\times10^5$ [41] |

## RESULTS AND DISCUSSION

### A. 2DEG concentration

Fig 4(a) shows the calculated 2DEG concentration as a function of the barrier thickness for N-polar AlN on β-Ga$_2$O$_3$ and metal-polar AlN on GaN. The polarization-induced 2DEG charge density which comes from donor-like surface states can be tuned by changing the barrier thickness. As the barrier thickness increases, the 2DEG concentration will enlarge and approach the polarization sheet charge density on the interface until the strained barrier relaxes.

It has been experimentally demonstrated that the AlN barrier thickness for 2DEG formation in metal-polar AlN/GaN HEMT is around 1.3nm[45]. It also can be estimated that the critical thickness of AlN barrier on GaN is ~6.5nm when taking Blanc's estimation $t_{cr} \sim b_e/2\epsilon$ (ε is the strain in AlN layer coherently strained on GaN, $b_e$ is the length of the Burgers vector)[46]. Therefore, a high-mobility window for AlN barrier thickness between 3~5nm (AlN is assumed to be coherently strained on GaN) with the sheet densities from ~$2.5\times10^{13}$/cm$^2$ to ~$3.9\times10^{13}$/cm$^2$ at AlN/GaN heterojunctions can be achieved[47]. Some other researches also account for it[48][49][50]. Our calculation results of metal-polar AlN/GaN HEMT based on H reference shows a good match with the experimental data in Fig 4 (a). Typically, for AlN barrier equals to 3nm, a 2DEG concentration confined at the heterointerface of $2.6\times10^{13}$/cm$^2$ can be achieved. Not only that, the designed N-polar AlN/β-Ga$_2$O$_3$ HEMT which holds a large sheet charge density (1.452 C/m$^2$) can induce a high electron concentration (~$10^{14}$/cm$^2$) that forms a 2DEG channel on the side of β-Ga$_2$O$_3$ as shown in Fig 4 (a). Even the barrier thickness is reduced to one monolayer like 3 Å, it can still hold a large 2DEG density($2.5\times10^{14}$/cm$^2$). As the thickness of barrier increases, the 2DEG concentration will enlarge and reach saturation at around 3nm (corresponding to $5.1\times10^{14}$/cm$^2$ 2DEG density).

Normally in Al(Ga)N/GaN HEMT using polarization-induced 2DEG, the 2DEG

concentration is very sensitive with the barrier thickness. Especially for metal-polar AlN/GaN HEMT, owing to the spontaneous polarization of AlN (1.333 C/m$^2$) and GaN (1.339 C/m$^2$) are very similar, upon reaching above the critical thickness (~6.5nm), the AlN barrier will be fully relaxed on GaN buffer layer and the 2DEG will disappear[51]. Fig 4(b) shows the calculating 2DEG concentration versus the relaxation degree (we assume that 0% and 100% represent the barrier is fully strained and fully relaxed on the buffer layer, respectively) of the conventional AlN/GaN HEMT and designed Ga$_2$O$_3$-based HEMT. Table IV lists the corresponding sheet charge density $\Delta\sigma$ as a function of relaxation degree (from 0 to 100%) for these two structures. For conventional AlN/GaN HEMT, the metal-polar AlN bears large tensile strain on GaN and the piezoelectric polarization (-0.109C/m$^2$) makes the most contribution for the total sheet charge density (0.114C/m$^2$). As the relaxation degree increase, the tensile strain effect in barrier becomes weaken, which reduce the sheet charge density on the interface. As can be seen from Fig 4(b), after the relaxation degree reaches 60%, only 0.022×10$^{13}$/cm$^2$ 2DEG concentration can be obtained. With more strain relaxation, almost no 2DEG can be induced in the channel. N-polar AlN/β-Ga$_2$O$_3$ HEMT shows a different case that even the barrier is fully relaxed, it can still hole a large 2DEG concentration (4.7×10$^{14}$/cm$^2$) due to the SP dominating property that the large discontinuity of spontaneous polarization on the interface can provide a large sheet charge density for the induction of 2DEG.

Table IV The sheet charge density of metal-polar AlN/GaN and N-polar AlN/β-Ga$_2$O$_3$ as a function of relaxation degree.

|  | Sheet charge density ($\Delta\sigma$) C/m$^2$ | | | | | |
| --- | --- | --- | --- | --- | --- | --- |
| Relaxation degree (%) | 0 | 20 | 40 | 60 | 80 | 100 |
| Metal-polar AlN/GaN | 0.114 | 0.092 | 0.071 | 0.049 | 0.027 | 0.006 |
| N-polar AlN/β-Ga$_2$O$_3$ | 1.452 | 1.429 | 1.405 | 1.381 | 1.357 | 1.333 |

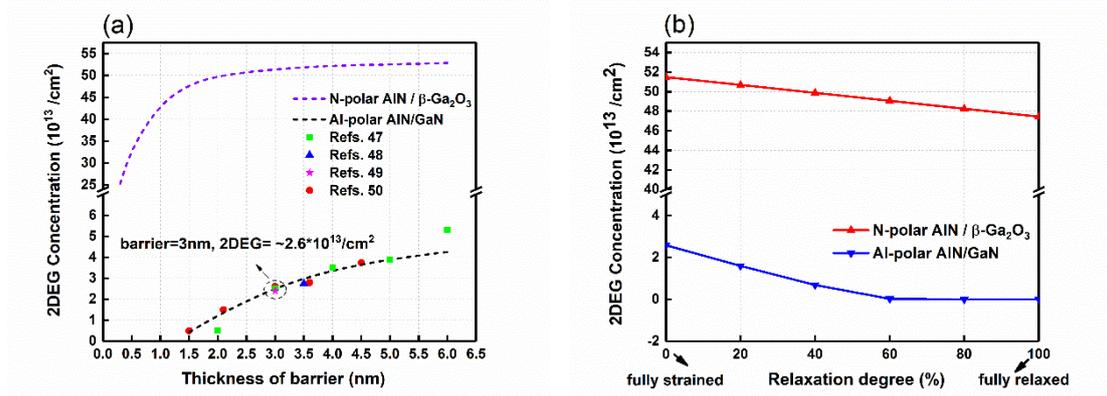

Figure 4. (a) Calculating 2DEG concentration as a function of thickness of barrier for N-polar AlN/β-Ga$_2$O$_3$ HEMT (violet dash line) and metal-polar AlN/GaN HEMT (black dash line) through H reference model. The different color symbols are taken from reference papers (Refs. 1-4) on metal AlN/GaN. The calculation results by H reference model show good consistent with the experimental data. (b) 2DEG concentration versus difference relaxation degree for N-polar AlN/β-Ga$_2$O$_3$ HEMT and metal-polar AlN/GaN HEMT through H reference model. Both structures employ 3nm barrier as an example.

### B. Output characteristics

Through the utilization of polarization engineering, a large 2DEG density on a HEMT device based on N-polar AlN/β-Ga2O3 heterostructure can be realized. For future power electronics applications, the DC common source family and transfer characteristics are also needed, as shown in Fig 5(a), (b). The proposed HEMT device shows a high drain saturation current ($I_{d,\,sat}$) of 18.5A/mm at $V_{DS}$=50V, $V_{GS}$=-5V, compared with 1.5~2.3A/mm in a previously reported AlN/GaN HEMT[45,46,47]. Thanks to the huge 2DEG concentration in the channel, it shows a very low on-state resistance $R_{on}$ of 0.28 ohm·mm, indicating a potential lower conduction loss. The transfer characteristics of the device demonstrate a much higher threshold voltage of -53.2V, corresponding to the normally-on operation mode. And the transconductance $g_m$ reaches its peak value of 1218mS/mm at $V_{GS}$=-49.7V, $V_{DS}$=10V. The transconductance is much larger than the conventional metal polar AlN/GaN HEMTs, which are reported to be between 300~480 mS/mm.[46,47,48] The large source-drain current and high transconductance are attributed to the huge 2DEG concentration on the interface.

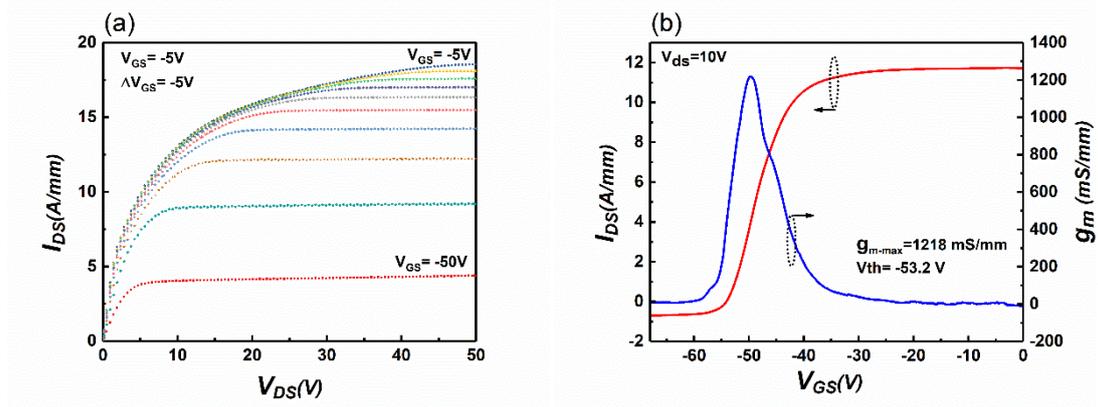

Figure 5. (a) Output characteristics measured with gate bias $V_{GS}$ from -5V to -50V at a step of -5V. (b) Transfer characteristics measured under a drain bias of $V_{DG}$=10V. The gate length, gate-drain spacing, and source-drain spacing of the device are $L_G$=1μm, $L_{GD}$=10μm, and $L_{SD}$=13μm, respectively.

## C. Reverse characteristics

Another feature of the proposed HEMT is the breakdown voltage of the channel layer. Fig 6 shows the breakdown voltage ($V_{BR}$) values of the proposed N-polar AlN/β-$Ga_2O_3$ HEMT. The $V_{BR}$ is defined at a drain leakage current of 1 mA/mm[52] while the gate voltage is applied to pinch-off the device before the drain voltage was increased until breakdown. It has been reported that GaN-based HEMT with the same gate to drain spacing (drift region) $L_{GD}$ as well as the same other settings on device structure, can only reach 560V with $L_{GD}$ = 10μm. From Fig 6, it can be easily observed that the breakdown voltage of proposed N-polar AlN/β-$Ga_2O_3$ HEMT reaches as high as 998.4V without any optimization such as the introduction of the localized Mg-doped layer under the 2-DEG channel[53] or the drain metal extension[54]. The enhancement of $V_{BR}$ is mainly due to the larger critical electrical field of β-$Ga_2O_3$ channel than GaN as the channel layer in conventional Al(Ga)N/GaN HEMT, and thus together with high current density, the proposed HEMT will potentially have a great power density performance. In addition, some optimization methods including the introduction of field plate, guard rings, as well as floating metal rings can be taken to further improve the breakdown voltage of $Ga_2O_3$-based HEMT.

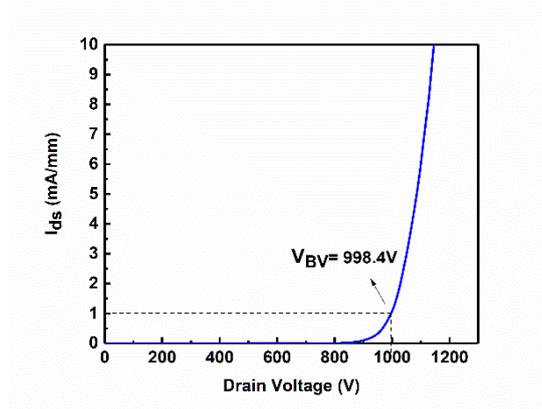

Figure 6. Off-state characteristics for the large area device of N-polar AlN/Ga$_2$O$_3$ HEMT with L$_{GD}$ = 10μm.

**CONCLUSIONS**

To fully excavate the potential of Ga$_2$O$_3$ on power device, especially HEMT, we propose the N-polar AlN/β-Ga$_2$O$_3$ HEMT by shifting the paradigm from δ-doping-induced to polarization-induced 2DEG. By employing the N-polar III-N materials on β-Ga$_2$O$_3$, the positive polarization-induced electric field along the growth direction in barrier can be created, which provides driving force for the electron into the channel. Further study shows that only AlN among all III-N materials can form the 2DEG channel on β-Ga$_2$O$_3$ due to the conduction band difference. In addition, the proposed N-polar AlN/β-Ga$_2$O$_3$ HEMT exhibits the SP-dominated sheet charge, resulting in a huge 2DEG concentration in the channel even with an ultrathin barrier. The large 2DEG concentration also contributes a lot to the output characteristics including high source-drain current, low on-state resistance, and high transconductance. High breakdown voltage is achieved due to the high critical field of β-Ga$_2$O$_3$ compared with GaN or SiC. This straightforward and effective paradigm can inspire the Ga$_2$O$_3$ and HEMT community to incorporate the polarization effect of III-N materials/ Ga$_2$O$_3$ interface on the design of high performance HEMT.

**AUTHOR INFORMATION**

*Equal contribution

Corresponding Author

#xiaohang.li@kaust.edu.sa

Notes

The authors declare no competing financial interest.

## ACKNOWLEDGEMENTS

The KAUST authors acknowledge the financial support from KAUST Baseline BAS/1/1664-01-01, KAUST CRG URF/1/3437-01-01, GCC REP/1/3189-01-01, and National Natural Science Foundation of China (Grant No.61774065). The work at Institute of Semiconductors, Chinese Academy of Sciences was supported by the National Key R&D Program of China (Nos. 2016YFB0400803 and 2016YFB0400802).